\documentclass[12pt]{article}

\usepackage{amsfonts}
\usepackage{amssymb}
\usepackage{graphics}
\usepackage{epsfig}
\usepackage{multirow}

\usepackage{subfig}
\usepackage{latexsym}
\usepackage{amsmath}

\usepackage{relsize}
\usepackage{geometry}
\usepackage{color}

\newcommand{\beq}{\begin{equation}}
\newcommand{\eeq}{\end{equation}}
\newcommand{\beqn}{\begin{eqnarray}}
\newcommand{\eeqn}{\end{eqnarray}}
\newcommand{\bea}{\begin{eqnarray}}
\newcommand{\eea}{\end{eqnarray}}
\newcommand{\beas}{\begin{eqnarray*}}
\newcommand{\eeas}{\end{eqnarray*}}

\newcommand{\bquo}{\begin{quote}}
\newcommand{\enqu}{\end{quote}}

\newcommand{\gsim}{\lower.7ex\hbox{$\;\stackrel{\textstyle>}{\sim}\;$}}
\newcommand{\lsim}{\lower.7ex\hbox{$\;\stackrel{\textstyle<}{\sim}\;$}}

\def\stroke{\vrule height8pt width0.4pt depth-0.1pt}
\def\topfleck{\vrule height8pt width0.5pt depth-5.9pt}
\def\botfleck{\vrule height2pt width0.5pt depth0.1pt}
\def\Zmath{\vcenter{\hbox{\numbers\rlap{\rlap{Z}\kern 0.8pt\topfleck}\kern
2.2pt
                   \rlap Z\kern 6pt\botfleck\kern 1pt}}}
\def\Qmath{\vcenter{\hbox{\upright\rlap{\rlap{Q}\kern
                   3.8pt\stroke}\phantom{Q}}}}
\def\Nmath{\vcenter{\hbox{\upright\rlap{I}\kern 1.7pt N}}}
\def\Cmath{\vcenter{\hbox{\upright\rlap{\rlap{C}\kern
                   3.8pt\stroke}\phantom{C}}}}
\def\Rmath{\vcenter{\hbox{\upright\rlap{I}\kern 1.7pt R}}}
\def\Z{\ifmmode\Zmath\else$\Zmath$\fi}
\def\Q{\ifmmode\Qmath\else$\Qmath$\fi}
\def\N{\ifmmode\Nmath\else$\Nmath$\fi}
\def\C{\ifmmode\Cmath\else$\Cmath$\fi}
\def\R{\ifmmode\Rmath\else$\Rmath$\fi}

\def\2{{1\over 2}}

\def\4N{${\cal N}=4$}
\def\N{{\mathcal N}}

\def\beq{\begin{equation}}
\def\eeq{\end{equation}}
\def\ba{\beq\new\begin{array}{c}}
\def\ea{\end{array}\eeq}

\def\R{{\rm R}}

\def\Z{\mathrm Z}
\def\1{\mathbbm{1}}
\def\N{{\cal N}}

\def\C{\rm C}

\def\1{\mbox{\tiny (1) }}
\def\0{\mbox{\tiny (0) }}

\begin{document}

\begin{titlepage}

\begin{flushright}
DAMTP-2010-46 \\
October 2010
\end{flushright}

\vspace{1mm}

\begin{center}
{\large  {\bf Magnetic Bags and Black Holes  }}
\end{center}

\vspace{0.1mm}

\begin{center}
{\large   {\large S. {\sc Bolognesi} } }

\vspace{0.1mm}

  {\it \footnotesize  DAMTP, 
Centre for Mathematical Sciences, \\
Wilberforce Road, 
Cambridge, CB3OWA, UK }

\end{center}

\begin{abstract}

We discuss gravitational magnetic bags, i.e., clusters of large numbers of monopoles in the presence of gravitational effects.
Physics depends on the dimensionless ratio between the vev of the Higgs field at infinity and the Planck mass. 
We solve the equations for the gravitational bags, and study the transition from monopole to black hole. 
The critical coupling for this transition is $v_{{\rm cr}} = \sqrt{\pi}/(4 \sqrt{G})$, and it is larger than that of a single 't Hooft-Polyakov monopole.
We investigate the black hole limit in detail.

\end{abstract}

\end{titlepage}

\vfill\eject

\section{Introduction}

Magnetic monopoles coupled to gravity have been the subject of constant developments over the last few decades.
Most of the work has been focused on a single 't Hooft-Polyakov monopole and the transition from regular space-time to a black hole with the horizon.
In this paper we will study the case of a large number of monopoles.
When $n$ is large, and the monopoles are sufficiently close to each other, the  solution can be approximated by a magnetic bag \cite{Bolognesi:2005rk} (see \cite{Lee:2008ze,Ward:2006wt} for related developments).
Magnetic bags are a closed surface, with a generic shape but constrained size, inside which the Higgs field is essentially zero, and outside which there are two free fields: the Abelian magnetic field and the massless scalar Higgs field. 
We are interested in finding equations and  solutions for the magnetic bag coupled to gravity.

The theory we consider is Einstein-Yang-Mills-Higgs (EYMH), SU$(2)$ Yang-Mills plus one Higgs adjoint field coupled to gravity: 
\beq
S_{\rm EYMH}= \int d^4 x \sqrt{-g} \left[ -\frac{1}{16 \pi G } R - \frac{1}{4 g^2} (F_{\mu\nu}^a)^2 + \frac{1}{2} (D_{\mu}\phi^a)^2  \right] \ .
\eeq
We work in the units $\hbar=c=1$ and $G =M_{{\rm Pl}}^{-2}$ and restrict to the vanishing  potential $V(\phi)=0$.
We also set the coupling $g=1$ for simplicity, since it does not enter in the relation between monopoles and black holes.
We call $v$ the value of $\langle \phi \rangle$ at infinity. The dimensionless coupling 
\beq
\alpha = \frac{v}{M_{{\rm Pl}}} 
\eeq
is the ratio between the Higgs vev and the Planck mass, and determines how much gravity is relevant in the monopole solution.
For small $\alpha$, the gravitational back-reaction is negligible; for higher $\alpha$ it is instead important. At a certain critical value $\alpha_{\rm cr}$ the monopole becomes so heavy and concentrated that it deforms the space-time around into a black hole.

There are at least three good reasons to consider    gravity and  large-$n$ monopoles together:
\begin{itemize}
\item It simplifies  the computations. The gravitational bag is simpler than the gravitating 't Hooft-Polyakov monopole, and more can be extracted from the solutions. We will use very simple numerics, and no shooting of parameters is required.
\item  There is binding energy between monopoles at finite $\alpha$. So it is natural for them to form stable agglomerates. 
\item The transition to black holes happens in the region where gravity is semi-classical. For a single monopole this couldbe achieved only by sending the coupling constant $g$ to zero. For large-$n$ monopoles this is not necessary.
\end{itemize}

The paper is divided into three main sections. In Section \ref{first} we review the magnetic bags, and discuss the low-energy dynamics. In Section \ref{second} we discuss the gravitating magnetic bags. 
Here we find an easy way to solve the differential equations numerically.
Much information can be extracted from the solutions. 
We also investigate in detail the black hole limit and how the various parameters behave near the critical coupling. In Section \ref{third} we conclude by discussing some open problems.

\section{Monopole Bags} \label{first}

In this section we consider the case without gravitational effects ($\alpha \to 0$).
The basic idea behind magnetic bags \cite{Bolognesi:2005rk}  goes as follows.
Consider the cluster of $n$ monopoles with $n \gg 1$. We consider the monopoles very close to each other (in a way that will be quantified later), so that their effect at a large distance is just that due to a source for the massless fields.

The boundary condition is $\phi^a = v \delta^{3a}$, and so the massless scalar fields are just a $U(1)$ gauge $A_{\mu}^3$ and the massless Higgs field $\phi^3$.
The monopole is specified by a  homotopic map from the sphere $S^2$ at spatial infinity the sphere $S^2$ of the vacuum manifold. We consider here a gauge in which this winding numbers  disappear, apart from leaving a Dirac string somewhere.

Since the electromagnetic field is free outside the bag (charged $W_{\mu}$ bosons have mass much higher than the scale we are considering now), we can introduce a magnetic potential $\varphi$ so that the magnetic field is the gradient of it, $\vec{B}= \vec{\nabla} \varphi$. Maxwell equations are then satisfied if $\varphi$ is a harmonic function $\triangle \varphi =0$.
The Higgs field $\phi$ is also free and obeys the Laplace equation $\triangle \phi =0$.
The non-trivial information is encoded in the boundary conditions at $r \to \infty$ and on the bag surface.

Let's work out the simplest bag, a sphere of radius $R$.\footnote{It is known that spherical symmetric monopoles do not exist for $n$ higher than $1$ \footnote{Weinberg:1976eq}. But for us spherical symmetry is achieved as an average in the limit of large numbers of monopoles (see \cite{Lee:2008ze}). In other words, spherical symmetry is always broken by $1/n$ corrections.}
Outside the sphere there are two scalar fields, the Higgs field $\phi$ and the magnetic scalar potential $\varphi$.
Both fields are harmonic (they satisfy the Laplace equation), but with different boundary conditions. 
$\phi$ has value $0$ on the bag surface and $v$ at infinity (Dirichlet boundary conditions);
$\varphi$ has Newman boundary conditions for the zero mode, since the total magnetic flux $4\pi r^2 \partial_r \varphi$ must remain constant and equal to $4 \pi n$. Is like solving the electrostatic problem for a conductor, first at fixed potential and then at fixed charge.
Their solution, as a function of $R$, is then  
\beq
 \varphi(r)= - \frac{n}{r} \ , \qquad \phi(r)=v\left( 1 - \frac{R}{r} \right) \ .
\eeq
The radius of the bag $R$ is fixed at the end by the energy, or mass,  minimization:
\bea
M(R) &=& \int_R^{\infty} 4\pi r dr \left( \frac{|\vec{\nabla} \varphi|}{2} + \frac{|\vec{\nabla} \phi|}{2} \right) \ , \\
&=&  2\pi \left( \frac{n^2}{R} +  v^2 R \right) \ ,
\eea
which gives as a result the radius $R$ and the mass of the monopole bag 
\beq
M= 4\pi n v \ , \qquad R_{\rm b}= \frac{n}{v} \ .
\eeq

Since the fields $\vec{B}$ and $\phi$ outside the bag already saturate the Bogomolny bound, we must conclude that the interior has no energy at all and the Higgs field vanishes.
This effect is similar to what happens for vortices \cite{Bolognesi:2005zr}.
For the monopole wall there is a very similar effect
of the flattening of the Higgs field on one side of the wall \cite{Ward:2006wt}.

The spherical shape is not the only one possible. In fact, every shape can lead to a magnetic bag with the same mass, the only constraint is on  its  size.
Let's prove that. Take a center point $(0,0,0)$, and a generic shape $S_{\lambda}$ where $\lambda$ is a scaling parameter from the center. We want to show that in this sequence there is one and only one bag that saturates the Bogomolny bound, and thus has the same mass of the spherical bag.
Given a surface $S_{\lambda}$, we need to solve the Laplace equation for $\phi$ and $\varphi$ with the boundary condition
$\phi_S =0$, $\phi_{\infty}=v$ and $\varphi_S={\rm const}$, $\varphi_{\infty}= -n/r + {\cal O}(1/r^2)$. 
\beq
\label{scaling}
\phi(r) \to  \phi(\lambda r) \qquad \varphi(r) \to \lambda \varphi(\lambda r)
\eeq
Whatever  the shape, is easy to see that the contribution of $\phi$ to the mass scales like $\lambda^{-1}$ while the contribution of $\varphi$ scales like $\lambda$. 
We have the mass
\bea
M(S_{\lambda}) &=& \int_{{\rm out}\, S_{\lambda}} d^3r   \ \left( \frac{|\vec{\nabla} \varphi|}{2} + \frac{|\vec{\nabla} \phi|}{2} \right) \ , \nonumber \\
&=&  4 \pi n v + 
\int_{{\rm out} \, S_{\lambda}} d^3r    \  \frac{1}{2} \left( \vec{\nabla} \varphi - \vec{\nabla} \phi \right)^2 ,
\eea
where the condition ${\rm out}\, S_{\lambda}$ means that the integral is restricted only to the portion of space outside the bag.
In the last line we have performed the Bogomolny decomposition. 
The remaining term is $\vec{\nabla} \varphi \vec{\nabla} \phi =  \vec{\nabla} ( \phi \vec{\nabla} \varphi )$ and  can be evaluated as a surface integral at infinity  $\int_{ r \to \infty } \phi \vec{\nabla} \varphi \cdot \vec {ds} =
4 \pi   n v $ and its value does not depend on $\lambda$.
A minimum of the mass is thus achieved for a certain value $\lambda^*$ when the Bogomolny equation $\vec{\nabla} \varphi = \vec{\nabla} \phi$ is satisfied, which is equivalent to $\phi = v + \varphi$.  The scaling (\ref{scaling}) ensures that such a $\lambda^*$ exists and is unique.
And thus the bag $S_{\lambda^*}$ also saturates the Bogomolny bound.

For axial symmetric elliptical shapes we can solve the problem.
We take an elliptic bag defined by the equation
\beq
\frac{x^2}{a^2} + \frac{y^2}{b^2} +\frac{z^2}{c^2} =1 \ .
\eeq
For any given ratio $a:b:c$, there is a unique overall size that satisfies the bag requirements.
The solution of the Laplace outside the ellipse is known (see \cite{Landau} or \cite{kel} for the detailed formulae)
The condition to impose is that the potential is $\varphi_0 = -v$ on the bag surface.
For axial symmetric ellipses with $a=b$ we have the following constraint:
\bea
&& \frac{v}{n} = \frac{\arccos (c/a)}{\sqrt{a^2 - c^2}} \qquad \ a=b \geq c  \ ,\\
&& \frac{v}{n} =  \frac{ {\rm arccosh} (c/a)}{\sqrt{c^2 - a^2}} \qquad a=b \leq c  \ .
\eea
These equations define a continuous one-parameter family of magnetic bags. For every $c$ there is a related $a=b$  uniquely defined by the parameters $n$ and $v$. The spherical bag with radius $n/v$ belongs to this family.
Another surface is the magnetic disk of radius $a=b= \pi n/(2 v)$ and $c=0$.
We used the exact solution for generic $n$ axially \cite{ Ward:1981jb} symmetric monopoles and verify that in the large-$n$ limit they approach the magnetic disc \cite{Bolognesi:2005rk}.

\begin{figure}[h!t]
\begin{center}
\leavevmode
\epsfxsize 6 cm
\epsffile{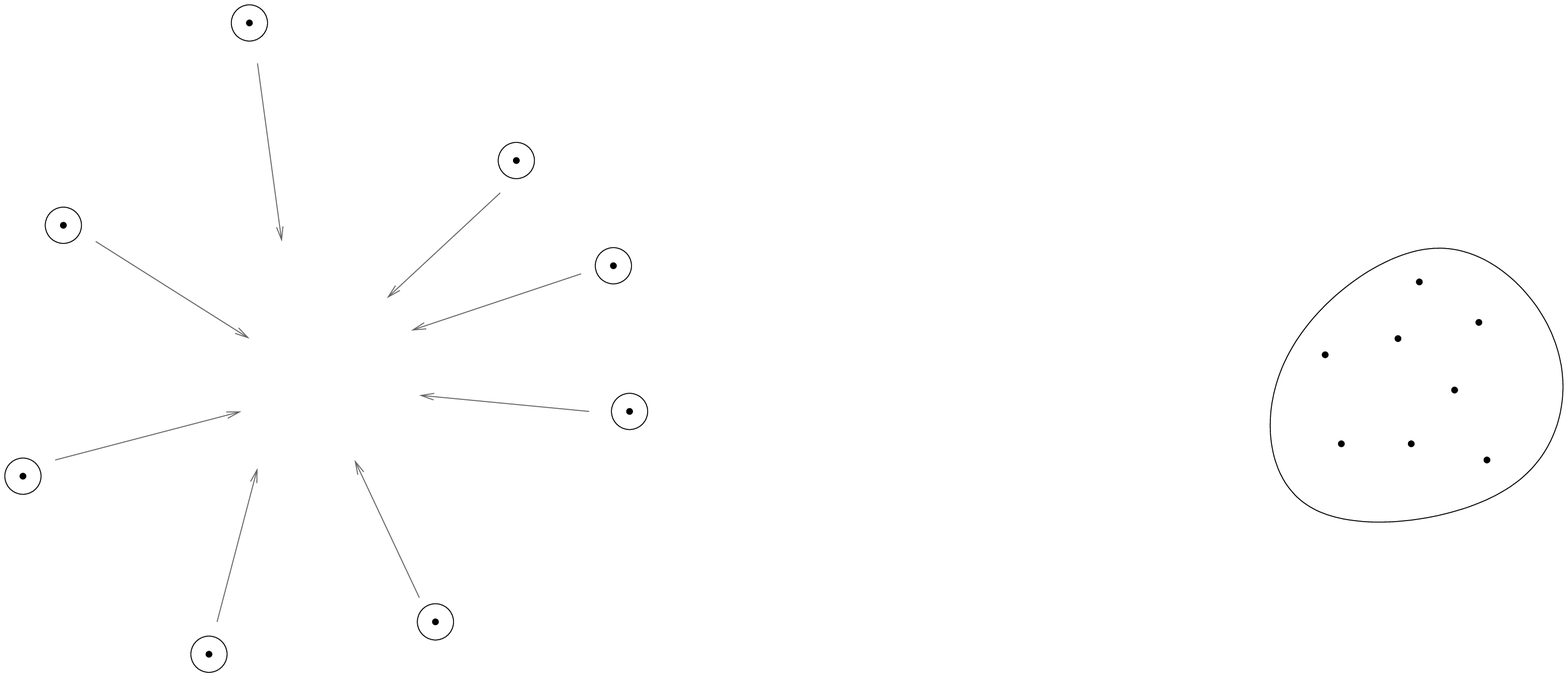}
\end{center}
\caption{\footnotesize Formation of the bag when monopoles are enough close to each other.  }
\label{analogy}
\end{figure}

In \cite{Lee:2008ze}, the authors showed that a bag does not uniquely specify a configuration in the monopole moduli space. 
For example if we take single monopoles far apart, and we make them come close to each other gradually (Fig. \ref{analogy}),  the bag surface is created when their mutual distance reaches the bag scale.
Clearly we can still move the zeros of the Higgs field inside the bag, without changing its shape.
But any motion of the zeros that does not change the bag surface  is completely detached from the energy.

For example we can ask what is the effective metric that a probe monopole monopole feels in the background of a large bag. 
We want to consider the moduli of the bag as fixed, and just compute the metric on the four coordinates of the monopole. We are not thinking about the gravitational metric but the one of the moduli space.
The kinetic  term is proportional to the Higgs vev and thus decreases as the monopole becomes closer to the bag.
In some sense this is the opposite limit of the one considered in \cite{Gibbons:1995yw,Bielawski:2007tq} where the metric of the moduli space for distant monopoles was studied. 

Even without gravity effects ($\alpha \to 0$), we find some properties of magnetic bags that are reminiscent of black holes.
First of all, note that the radius of the bag scales like $n$. 
This is certainly not typical of compact objects, where the radius scales like $n^{1/3}$. 
This is instead like  black hole's horizon, which scales like the total mass.
An obvious difference instead is that the surface of the magnetic bag can fluctuate.
Note also the bag, from the energetic point of view, is essentially a {\it hole}; there is no energy inside. 
If a probe monopole falls into a magnetic bag, it loses its identity. 
The monopole   contribution to the entropy can only be in changing the shape of the bag surface.
In the next section we add gravity and interpolate between the magnetic bag and the black hole.

We conclude the flat space-time section  with a discussion about the metric in the moduli space of magnetic bags.
The problem is not easy to solve explicitly for one reason. 
There are no obvious coordinates in which both the metric and the constraint are easily implementable.
For example we can choose to parametrize the bag with the harmonic function  $\varphi $ at infinity.
We then have an infinite dimensional space that parameterizes the bag surface one to one. 
The constraint on the size is simple, it is  just that the coefficient of the $1/r$ term of the harmonic function  is fixed to be $-1$ (we consider $n=1$ now). 
But for the metric we have to solve the inverse problem, from the harmonic function to the surface $\varphi = -v$.

We can choose to parameterize with the harmonic function $\varphi$ that implicitly defines the magnetic bag; $S_{\varphi}$ is defined as the equisurface where $\varphi = -v$.
We have to perturb the bag, which is for us a perturbation of the harmonic function  $\varphi + \delta \varphi(t)$, and insert this into the action.
First we need the zero mode condition; it is just that $\delta \varphi = {\cal O}(1/r^2)$, and does not have any $1/r$ term.
Then the deformation of the  surface of the bag $S_{\varphi+ \delta \varphi} $ which is  obtained by the 
\beq
\vec{x} \to \vec{x} + \frac{\vec{\nabla}\varphi \  \delta \varphi}{|\vec{\nabla}\varphi|^2} \ ,
\eeq
where $\vec{x}$ is any point on the surface $S_{\varphi}$.
The metric of the moduli space is given by the norm of the zero mode; there are two terms to take into account. One is simply the norm of the Higgs field zero mode outside the surface, which is
\beq
\int_{{\rm out} S_{\varphi}} \frac{(\delta \varphi)^2 }{2}
\label{firstzero}
\eeq
If we consider the simplest case of a spherical bag in straight motion, the zero mode is a dipole term 
\beq
\label{deltaphi}
\delta \varphi = \frac{\vec{r} \ \vec{n} }{r^3} \ ,
\eeq
where $\vec{n}$ is the direction of motion.
The norm gives
\bea
S 
&=& 
\left( \int_{{\rm out } \  S_{\varphi}} d^3 x \frac{\cos^2 \theta}{r^4}    \right) \ \int dt  \frac{\dot{\vec{n}}^2}{2}  \label{somma} \\
&=& \frac{4 \pi v}{3}  \int dt  \frac{\dot{\vec{n}}^2}{2}
\eea
This gives only one third of the required mass $4 \pi v$. The other two thirds comes from the electric field.
A magnetic charge in motion, such as the magnetic bag, generates electric circuiting around the line of motion.
The spherical bag in straight motion generates the electric field
\beq
\vec{E} = \frac{\vec{r} \wedge \dot{\vec{n}} }{r^3} \ ,
\eeq
which is maximal on the equator and zero on the poles, the opposite of the Higgs field zero mode. The kinetic term for the electric field gives the missing two thirds in the mass.




\section{Gravitating Magnetic Bags} \label{second}

We want to study the magnetic bags for generic coupling $\alpha$.
For a spherically symmetric configuration, a generic metric can be recast in the form
\beq
ds^2 = B(r) dt^2 - A(r)^{-1} dr^2 -r^2 d\Omega^2 \ .
\label{ansatzmetric}
 \eeq
A magnetic Reissner-Nordstrom black hole is always a solution of the equation of motion \cite{Bais:1975gu}.
In that case 
\beq
\label{rnbh}
B(r)=A(r)=1-\frac{2 G M}{r}+ \frac{4 \pi n^2 G}{r^2}
\eeq
with the extremal values
\beq
M=2n\sqrt{\frac{\pi}{G}} \ , \qquad  R_{{\rm h}}=2n \sqrt{\pi G} \ ,
\eeq
where $R_{{\rm h}}$ is the radius of the horizon.
Although the magnetic black hole is always a viable solution, for small $v$ is certainly lighter than a regular monopole, with no horizons. 
The existence of such solutions, for the single unit monopole, was first proved in \cite{VanNieuwenhuizen:1975tc}. Explicit numeric solutions were then found later
\cite{Ortiz:1991eu,Breitenlohner:1991aa,Lee:1991vy}.
When $\alpha$ reaches a maximal value $\alpha_{\rm max }$, the monopole becomes a black hole.
More recently multi-monopoles with axial symmetry have been studied \cite{Hartmann:2000gx} where in particular the authors have found that, for vanishing potential, multi-monopoles are gravitationally bound.

We shall focus here on the large $n$ magnetic bag limit.
The gravitational bag is a patched solution.
The surface of the bag is a co-dimensional one object that separates two domains of space-time.
Is empty  space-time inside $r<R$, and is magnetic field plus Higgs field tail outside $r>R$.
The inside of the bag is flat since is compact and there are no energy sources. The outside has two kind of sources for gravity, the Higgs field and the magnetic field. We then have to choose the proper way to patch the two space-times.

The action outside the bag can be written as
\bea
&& S_{{\rm out}} = - 4 \pi \int dt \int_R^{\infty} dr \left[ \frac{r}{8 \pi G} (1-A(r)) \left(\sqrt{\frac{B(r)}{A(r)}}\right)' \right. \nonumber \\
& & \qquad \qquad  + \left. r^2 \sqrt{\frac{B(r)}{A(r)}} \left( A(r) \frac{\phi'(r)^2}{2} + \frac{n^2}{2   r^4} \right) \right] \ ,
\label{action}
\eea
where the first term is the gravitational contribution and the second the matter contribution. 
The gravitational contribution differs from the Einstein-Hilbert by a total derivative term, as prescribed in \cite{VanNieuwenhuizen:1975tc}.
The unknown functions are the three profiles $A(r)$, $B(r)$ and $\phi(r)$.
The magnetic field is fixed and does not depend on the change in the metric. As long as the metric is expressed as (\ref{ansatzmetric}), the magnetic field is the same as in flat space.
The matter equation is
\beq
\partial_r (r^2 \sqrt{B(r) A(r)} \phi'(r)) = 0 \ .
\eeq
The equations for the metric fields are
\bea
 \frac{A(r)}{B(r)}\left(\frac{B(r)}{A(r)}\right)' &=& 8 \pi r G  \phi'(r)^2 
\label{firsteq}\\
r A'(r) -1 +A(r) + \frac{4 \pi G n^2}{  r^2}  &=& -4\pi G A(r) r^2 \phi'(r)^2 \ ,
\label{secondeq}
\eea
where the second is the extremization of the action with respect to $B(r)$ and the first by the combination $A[r] \frac{\delta S}{\delta A[r]} + B[r] \frac{\delta S}{\delta B[r]}$. 
The equations have been arranged so that is evident that without the contribution from $\phi'(r)0$ the solution is given by the magnetically charged black hole (\ref{rnbh}).

In terms of the Einstein equations $R_{\mu \nu} = -k S_{\mu \nu}$, the first is given by the sum $-r  R_{rr} -r R_{tt}/(A B) $ and the second by the sum $r^2 A R_{rr}/2 + R_{\theta \theta} + r^2 R_{tt} / (2 B)$. The reduced energy momentum tensor is $S_{\mu\nu} = T_{\mu\nu} -g_{\mu \nu} T_{\lambda}^{\lambda}/2 =  {\rm diag} [0,\phi'^2 ,0,0] + {\rm diag} [ B (n/r^2)^2/2,-(n/r^2)^2/(2 A),(n/r )^2/2 , \sin^2{\theta}(n/r )^2/2]$ where the first is the contribution from the scalar field and the second from the gauge field  (see \cite{Weinberg}).

The equation for the scalar field can be integrated 
\beq
\phi'(r) = \frac{a}{r^2 \sqrt{A(r)B(r)}} \ .
\eeq
The constant of integration $a$ will be fixed later. 
Inside the bag $r<R$ we have flat space  with no magnetic field, $\phi=0$ and 
\beq
A(r)=1  \ , \qquad B(r)= b \ .
\eeq
The constant $b$ is just a factor that can be obtained from Minkowsky space-time by rescaling the time coordinate $t \to t \sqrt{b}$. It is important to keep it free because we want to fix the metric $(1,-1,-1,-1)$ at $r \to \infty$ to be normalized to one. 
The factor $b$ is a physical quantity that measures the redshift from the bag surface  to the infinity. 
No such constant is possible for $A(r)$. In fact $A(R_{{\rm b}})=1$ is what defines the radius $R_{\rm b}$ from the solution of the equations (\ref{metriceq}).

To find a solution we start at $r \to \infty$ with the boundary 
\bea
A(r) &=& 1  -\frac{2 m_A}{r} +  \frac{q_A^2}{r^2} + \dots \\
B(r) &=& 1   -\frac{2 m_B}{r} +  \frac{q_B^2}{r^2} + \dots
\label{boundarycondition}
\eea
The $1/r$ terms are constrained to be equal $m_A =m_B =m$, but $m$ remains a free parameter at this stage. All the other coefficients are then fixed by the differential equations as the function of $m$.
We then have the equation (\ref{metriceq}) with boundary condition (\ref{boundarycondition}).  The solution is then uniquely specified by the choice of $a$ and $m$.

An important property of the equations, and  of the boundary conditions, is that the solutions for different $n$ and same $v$ are exactly the same, just rescaled $r \to n r$.
We can thus work with $n=1$ for the rest.
The boundary expansion also gives the scaling of the mass $m \to n m$.
This simple scaling property of the bag equation has an important, somehow counterintuitive, physical consequence. 
Gravitational bags still have mass vs. $n$ constant; the binding energy is a finite-$n$ effect.
It could be that the moduli space of shapes still persist even after the introduction of gravity, but for that we have no proof yet.

We choose a value for $a$ and $m$ and then simply solve numerically the differential equation.
The condition $A(R_{\rm b})=1$ then determines the radius of the bag $R$ where the monopole wall is located. 
The scalar field $\phi(r)$ can also be integrated. It starts from $\phi(R_{\rm b})=0$ and then saturates to a constant $\phi(\infty)=v$.
In this approach the vev $v$ appears at the end as function of $a$ and $m$.

Still we miss one last condition. 
Of the two free parameters $a$ and $m$, one should be fixed. It is the same to say that for a specific value of the boundary vev $v$ everything should be fixed. In this approach we start instead from $m$ and determine everything, $v$ included. 
The condition that fixes $a$ is the extremization of $v$ or the fact that $B(r)$ is patching with continuous derivative. 
We checked numerically that these two conditions are equivalent.

This is in fact an example of the Israel junction conditions. The bag surface is a thin time-like hypersurface that separates a flat and empty space-time from  one with magnetic field and scalar field excitation.  
The junction conditions says that the induce metric on the wall must be continuous $h_{i j}$. 
In our case the thin wall has no tension and no  dynamics localized on the wall.
So the other condition is just the continuity of the extrinsic curvature $K_{ij}=-\Gamma^r_{ij} = \partial_r g_{ij}/2$. 
These conditions for our case give previous patching conditions, namely $A(R_{\rm b})=1$ and $B(R_{\rm b})'=0$.

Having checked that the continuity of the derivative $B(R)'=0$ is the right patching condition, we can use it as the starting point of the computation to get a considerable simplification from the previous method.
The trick is to start not from $r \to \infty$ but from the bag surface $r = R_{\rm b}$.

The convenience of this method is that it does not require any minimization. Any solution obtained with a starting point $R \geq R_{{\rm cr}}$ is a good one. We just have to retrieve the corresponding $v(R)$ and $m(R)$.
We simply have to solve a second-order differential equation with regular boundary conditions at $r=R$.
\bea
\left(\frac{B(r)}{A(r)}\right)' &=& 8 \pi G  \frac{1}{r^3 A(r)^2}  \\
r A'(r) -1 +A(r) + \frac{4 \pi n^2}{r^2}  &=& -4\pi  G \frac{1}{r^2 B(r)}  
\label{metriceq}
\eea
with boundary conditions $A(R_{\rm b}) = B(R_{\rm b}) =1$. The coefficient $a$ has been fixed to be $1$ by the condition $B'(R)=0$. Now $B(\infty) = b^{-1}$.

\begin{figure}[h!t]
\begin{center}
\subfloat[$R_{{\rm b}}=6$ \ , $m=1.974..   $ ]{
\epsfxsize 6 cm
\epsffile{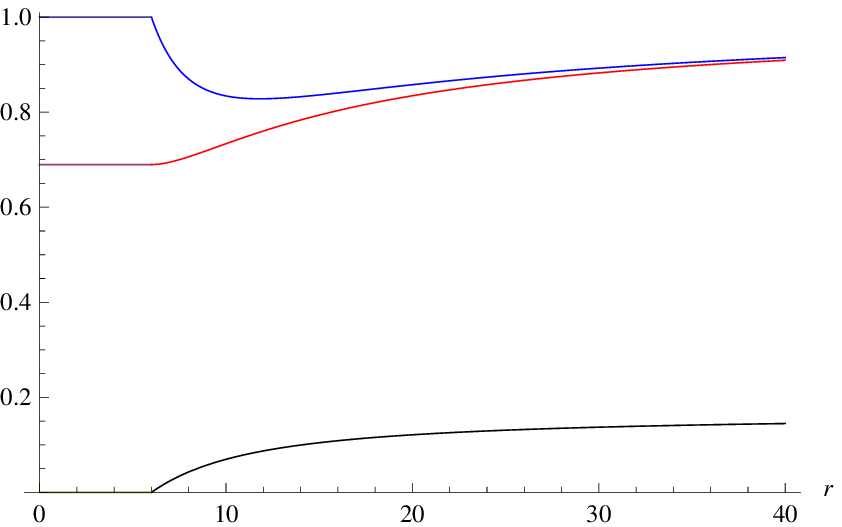}
}
\qquad
\subfloat[ $R_{{\rm b}}=4$ \ , $m=2.746..   $]{
\epsfxsize 6 cm
\epsffile{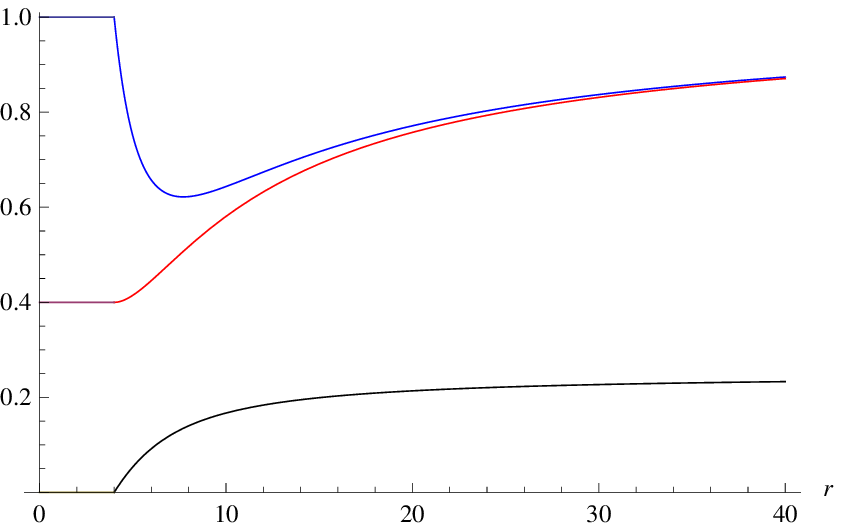}
}
\end{center}
\begin{center}
\subfloat[   $R_{{\rm b}}=2.6$\ , $m=3.468..  $]{
\epsfxsize 6 cm
\epsffile{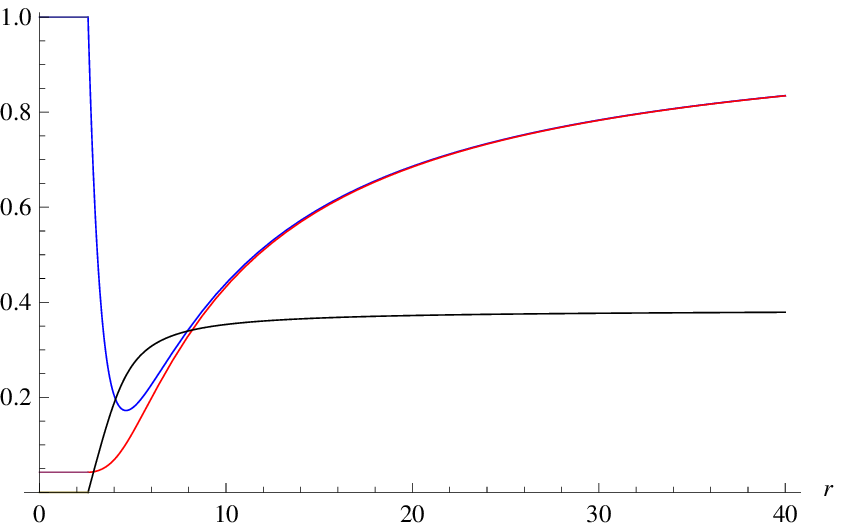}
} \qquad
\subfloat[  $R_{{\rm b}}=4/\sqrt{\pi}$\ , $m=2 \sqrt{\pi}  $ ]{
\epsfxsize 6 cm
\epsffile{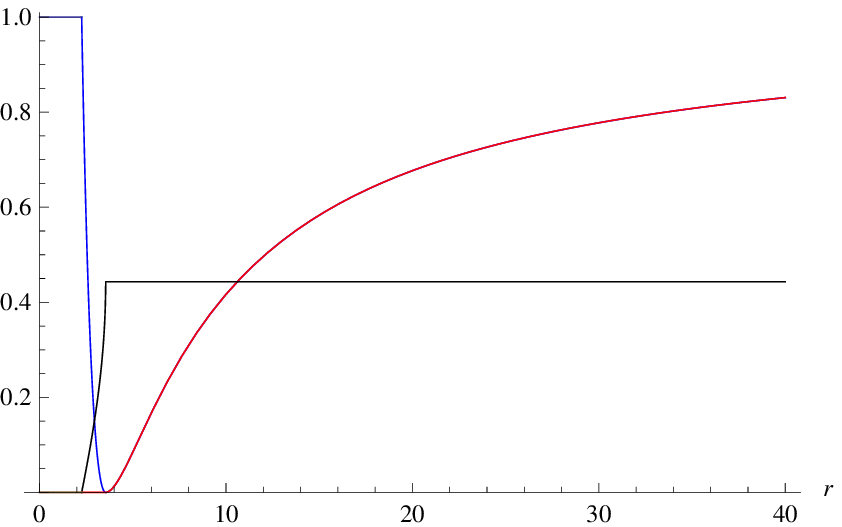}
}
\caption{\footnotesize Some solutions for various values of $R_{\rm b}$.}
\label{figura}
\end{center}
\end{figure}

\begin{figure}[h!t]
\begin{center}
\leavevmode
\epsfxsize 6 cm
\epsffile{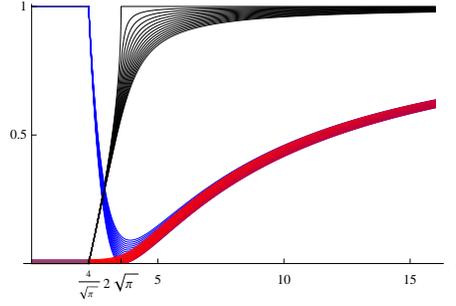}
\end{center}
\caption{\footnotesize A sequence of plots converging to the black hole (here the radius and $\phi$ have been normalized to $4/\sqrt{\pi}$ and $1$ for every $\alpha$).  }
\label{sequenza}
\end{figure}

We now present the result obtained. All the plots are obtained setting $n=1$ and in units $G=1$. In \ref{figura},
we have a sequence of profile functions for various values approaching the criticality. A rescaled sequence is given in \ref{sequenza}.


\begin{figure}[h!t]
\begin{center}
\leavevmode
\epsfxsize 6 cm
\epsffile{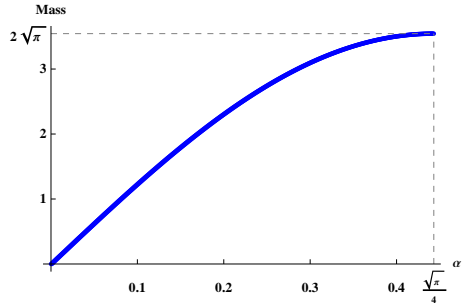}
\end{center}
\caption{\footnotesize Mass versus $\alpha$.   }
\label{mass}
\end{figure}

\begin{figure}[h!t]
\begin{center}
\leavevmode
\epsfxsize 6 cm
\epsffile{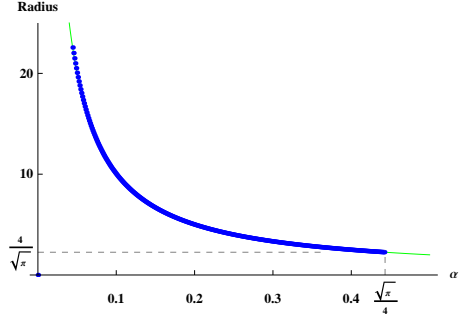}
\end{center}
\caption{\footnotesize Radius versus $\alpha$. Note that the plot is indistinguishable from $1/\alpha$.  }
\label{radius}
\end{figure}

\begin{figure}[h!t]
\begin{center}
\epsfxsize 6 cm
\epsffile{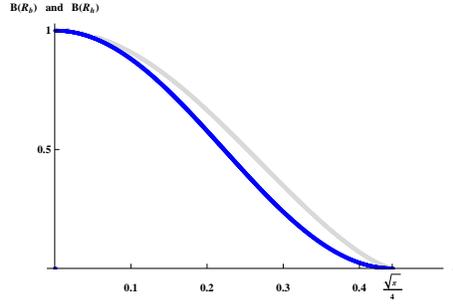}
\caption{\footnotesize The $B(R_{\rm b}))$ compared with $B(R_{\rm h})$. }
\label{bofr}
\end{center}
\end{figure}

In Figure \ref{mass} is the plot of the specific mass versus the coupling $\alpha$.
In Figure \ref{radius} is the plot of the radius as a function of $\alpha$.
The main features are as follows. At $\alpha \ll 1$ is linear with coefficient $4 \pi$. Then the attraction of gravity makes the pendence lower.
There is a critical value $\alpha_{\rm cr} = 4/\sqrt{\pi}$ where there is black hole formation and the specific mass saturates to the black hole value $2\sqrt{\pi/G}$.
The fact that the critical value is exactly $4/\sqrt{\pi}$ has been found numerically, and we do not have an analytic explanation for that. 
Note that the plot of the radius is indistinguishable from the function $1/\alpha$. This means that the flat space-time relation between $R$ and $v$ is {\it not} modified by gravitational correction at any order. 
We will later give an analytic proof for this using the functional approach. 
In Figure \ref{bofr} is the plot of the redshift factor $b$ versus the coupling $\alpha$.
This vanishes quadratically near the black hole limit. The redshift computed at the horizon vanishes only linearly.


In the limit $\alpha \to 0$ the  metric is   given by 
\bea
A(r) &=& 1  -\frac{2 G M}{r} +  \frac{4 \pi n^2 G}{r^2} + \dots  \nonumber \\
B(r) &=& 1   -\frac{2 G M}{r} +  \frac{8 \pi n^2 G}{r^2} + \dots
\label{}
\eea
with $M=4 \pi v n$. 
This is consistent with the weak-field limit approximation, in which the deviation from flat-space metric is very small.

The single monopole with gravity can be  defined as a minimum of a certain energy functional \cite{VanNieuwenhuizen:1975tc}.
The same is true for the magnetic bag. The steps to obtain this functional are exactly the same as the ones for the single 't Hooft-Polyakov monopole.
First we integrate (\ref{firsteq}) to obtain
\beq
\frac{B}{A} = \exp{\int_r^{\infty} ds \  8 \pi s G \phi'(s)^2} \ .
\eeq
We the insert this into the action (\ref{action})
\beq
S_{{\rm out}} = - 4 \pi \int dt \int_R^{\infty} dr   r^2   \left(   \frac{\phi'(r)^2}{2} + \frac{n^2}{2   r^4} \right) \exp{\int_r^{\infty} ds \  4 \pi s G \phi'(s)^2} \ .
\eeq
Note that also $A(r)$ disappears, without having used the information about its equation of motion (\ref{secondeq}).
This is a functional of both $R$ and $\phi(r)$.

We want to use the functional approach to compare the masses of the single monopole with those of the magnetic bag.
For the 't Hooft-Polyakov monopole taking the ansatz
\beq
\phi^a =  \hat{r}^a \phi(r) \qquad A^a_i = \epsilon_{iak} \hat{r}^k \frac{1-u(r)}{r}
\eeq
and the two functions
\beq K = \frac{u^{'2}}{r^2} + \frac{1}{2}  
\phi^{'2} \qquad U = \frac{(u^2-1)^2}{2r^4} + \frac{u^2 \phi^2 }{r^2} \ .
\eeq
Then the functional is
\beq
M_1[u(r),\phi(r)] = 4 \pi \int_0^{\infty} dr r^2 (K + U) \,  \exp{\left(-8 \pi \alpha^2 \int_r^{\infty} ds   s K\right) } \ .
\eeq
The formula  can be applied to the magnetic bag, with the only difference that now the kinetic and potential terms are given by
\beq
K =  \frac{1}{2} \phi^{'2} \qquad
U = \frac{1}{2r^4} \ .
\eeq
The functional to minimize to obtain the mass is then
\beq
M[R,\phi(r)] =  4 \pi \int_R^{\infty} dr r^2   \left( \frac{1}{2} \phi^{'}(r)^2 + \frac{1}{2r^4} \right)  \, \exp{\left(-4 \pi \alpha^2 \int_r^{\infty} ds   s \phi^{'}(s)^2\right)} \ ,
\label{functional}
\eeq
where both the parameter $R$ and the function $\phi(r)$ must be varied to find the minimum.

The functional gives important information. 
One obvious observation, for both functionals, is that for $\alpha \to \infty$ they  reduce to the usual energy functional. 
We know in that case that the two functionals have the same minimum, although reached with two different functions. For the 't Hooft-Polyakov it is the BPS solution
\beq
u(r)= \frac{r}{\sinh r}\ , \qquad \phi(r)= \frac{\cosh r}{\sinh r} -\frac{1}{r} \ ,
\eeq
while for the magnetic bag, it is  $R=1 $ and $\phi(r)=1-1/r$.
Another straightforward observation is that $\alpha$ decreases the minimum. For example taking the functions that minimize the functional for  a certain $\alpha_1$, and inserting them in the functional for $\alpha_2$, is already enough to lower the minimum. 
The mass vs. $\alpha$ plot must then be a convex function negative second derivative.
Making a direct comparison between the two functionals is less obvious because there are various competing effects. 
We do not expect  marginal stability at finite $n$, but we expect the bound
\beq
\label{comparison}
M_1 \geq M = \lim_{n \to \infty} \frac{M(n)}{n}
\eeq
with equality satisfied only in the limit $\alpha \to 0$.  
This expectation comes from the fact that forces between two elementary monopoles are, in total, attractive at a large distance.

\begin{figure}[h!t]
\begin{center}
\leavevmode
\epsfxsize 6 cm
\epsffile{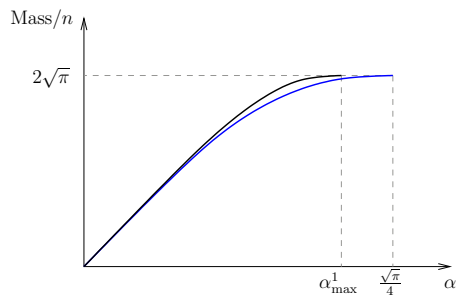}
\end{center}
\caption{\footnotesize  Mass versus $\alpha$ of the single monopole compared with that of the magnetic bag. According to \cite{Breitenlohner:1991aa} $\alpha^1_{\rm max} \simeq .39..$.}
\label{possibilities}
\end{figure}

From the energy functional, we can prove that this is the case for small $\alpha$.
Finding the first gravitational correction to the mass can be done with the only use of the $\alpha=0$ solution
and then inserting it directly into the first term of the exponential expansion \cite{VanNieuwenhuizen:1975tc}. This gives  
\beq
M_1(\alpha) =  4 \pi v (1 -1.844.. \alpha^2 + {\cal O}(\alpha^4) )
\eeq
Where $1.844..$ is a numerical evaluation of the integral so obtained.
The back-reaction of the profile function is of order $\alpha^2$, but this would give a correction of order $\alpha^4$ to the monopole mass.
The very same technique can be used for the magnetic bag.
Now the integral is simpler and can be computed exactly. The result is
\beq
M(\alpha) =  4 \pi v (1 -\frac{2 \pi}{3} \alpha^2 + {\cal O}(\alpha^4) ) \ .
\eeq
We thus directly prove that at first order $M$ is strictly smaller that $M_1$.
Although we expect this to be true for every value of $\alpha$, we do not have a rigorous proof for the bound (\ref{comparison}). 
We can for example compare the critical value in the literature for the single monopole, and the one obtained previously for the large-$n$ monopole.
 According to \cite{Breitenlohner:1991aa} $\alpha^1_{\rm max} \simeq .39..$ which is smaller that $\sqrt{\pi}/4 \simeq .44...$.
This seems to confirm a plot like Figure \ref{possibilities}, where the single monopole mass is always above the curve $M(\alpha)$ and reaches the black hole phase before the magnetic bag. 

The last information we can extract from the functional is a proof of non-renormalization of the relation between the radius of the bag and $v$ (Figure \ref{radius}).
The functional (\ref{functional}) can be rewritten to the rescaled coordinate  $x = r/R$. In this way $R$ disappears from the boundary of the integral, which is now fixed to be $1$ and $\infty$..
\beq
M[R,\phi(x)] =  4 \pi \int_1^{\infty} dx x^2   \left( \frac{1}{2} R \phi^{'}(x)^2 + \frac{1}{2 R^3 x^4} \right)  \, \exp{\left(-4 \pi \alpha^2 \int_x^{\infty} dy   y \phi^{'}(y)^2\right)}  \ .
\eeq
The two functionals are equivalent; still we need to make a variation with respect to $R$ and $\phi(x)$. 
The non-trivial thing is in the term inside the exponential {\it does not} contain $R$. This because the combination $ds s  \phi^{'}(s)^2$ is scaling invariant.
This term is the one that controls the dependence of the solution from the gravitational constant.
Changing $G$, but keeping fixed $v$, will thus alter only the profile of $\phi(r)$ but not the radius $R$.

\begin{figure}[h!t]
\begin{center}
\leavevmode
\epsfxsize 6 cm
\epsffile{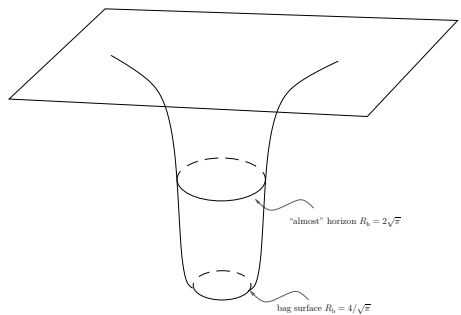}
\end{center}
\caption{\footnotesize Slice of the space geometry close to the critical coupling. The throat is not yet infinite, and is capped off by the magnetic bag.}
\label{geometry}
\end{figure}

The most interesting situation is when the geometry is close to be a black hole, but is not yet so.
There is an ``almost'' horizon at $R_{\rm h}= 2\sqrt{\pi}$, which corresponds to a long neck in the space geometry. A slice of the spatial geometry is like in Figure \ref{geometry}. 
The long neck is not infinite since it is capped off by the magnetic bag located at $R_{\rm b}=4/\sqrt{\pi}$.
We can make an expansion, around the critical value, of the various physical parameters. following
\bea
R&=&4/\sqrt{\pi} + \epsilon \ , \nonumber \\
v&=& 4/\sqrt{\pi} - \epsilon + {\cal O}(\epsilon^2) \ ,  \nonumber \\ 
M &=& 2 \sqrt{\pi} + M_{2}  \epsilon^2 + {\cal O}(\epsilon^3)  \ , \nonumber \\ 
b &=& b_{2} \epsilon^2 + {\cal O}(\epsilon^3)  \ .
\eea 
Where we choose the parameter $\epsilon$ as the deviation from the critical radius. So the first line is a conventional choice, the second we already know to all orders since $R=1/v$, and  the following are results for the mass the   redshift:
\beq
  M_{2}= 4 \pi^{3/2} \ , \qquad b_{2} = 4 \pi \ .
\eeq
As for  $\alpha_{{\rm cr}}$, we again find exact numbers for the coefficients of the first terms in the expansion.
These findings are also numeric.

\begin{figure}[h!t]
\begin{center}
\leavevmode
\epsfxsize 6 cm
\epsffile{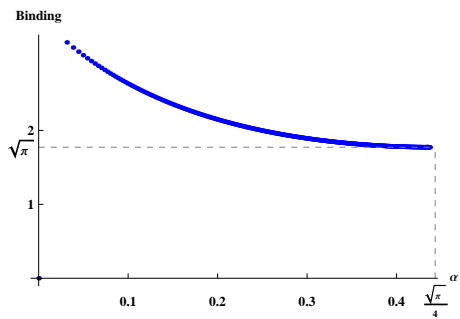}
\end{center}
\caption{\footnotesize Binding energy blue-shifted at the bag surface. }
\label{binding}
\end{figure}

In the window  $\alpha^1_{\rm max} < \alpha < \sqrt{\pi}/4$, where the single 't Hooft-Polyakov monopole is already a black hole but the magnetic bag is not, the discrepancy $ \delta = M - 2 \sqrt{\pi}-M=   4 \pi^{3/2} \epsilon ^2$  is the binding energy, that is the energy  required for a magnetic bag to enucleate a single monopole
\beq
M(n) + \delta = M(n-1) + M(1) \ .
\eeq
The binding energy vanishes as $\epsilon \to 0$; this is  because also the magnetic bag becomes a extremal RN black hole.
But this binding energy is computed from the point of view of an external observer, the flat space-time at $r \to \infty$.
It is instead interesting to compute what this binding energy is from the point of view of an observer in the flat space-time around the bag surface. We have to blue-shift multiplying by $1/B(R_{\rm b})$. 
The result is given in   plot \ref{binding}; the binding energy does not vanish in this frame but goes to $\sqrt{\pi}$.

This has an important consequence. Since there is binding energy, there are many exited states above the vacuum. All these existed states, from the point of view of an observer at infinity, are instead seen as degenerate (they are red-shifted by a factor $1/B(R)$).
We thus have   entropy in the monopole sector.

The argument for the existence of entropy is simple, although qualitative.
We can put the system in a thermal bath at temperature $T$. As long as the temperature is smaller than the binding energy, the single monopoles do not escape from the bag. Not we go to the critical limit $v \to v_{\rm cr}$.
The temperature, as measured from asymptotic infinity must vanish because it is bounded by the binding energy. From the point of view of the bag though it does not.
We thus conclude that some entropy, the degeneracy of states, must remain in the monopole sector when we reach the critical value of $v$.
These states are seen from the asymptotic observer as degenerate.
We know that the black hole has   entropy; we just have proved that part of it is in the multi-monopole moduli space.

For example a single monopole can create a black hole, and with $g \to 0$ even in semiclassical gravity, but the entropy is certainly not in the monopole moduli space, which is null \cite{Lue:2000nm}.
How to compute this monopole entropy, in order to compare it with the black hole entropy, remains an open problem.

\section{Conclusion} \label{third}

The main result presented in this paper is the solution for the magnetic bag coupled to gravity. 
Equations for the gravitating bag  are easier than the single 't Hooft-Polyakov ones. 
There are three profile functions $A(r), B(r), \phi(r)$.
With the trick of starting from the bag surface to solve the differential equations, we do not have any problem of mathing with boundary conditions at infinity.
Every solution we get is a good one; we just have to recover the value of $v$ and $m$ as function of $R$.
No shooting of boundary conditions is required.

Still  a number of hints   indicate that more analytical progress could be done. A curious fact emerges when we analyzed in detail the near black hole limit.
The critical value of $v$ appears to be an exact number, namely $\sqrt{\pi}/4$. This can be checked empirically with great precision.
In addition the expansion of the other parameters near the black hole limit seems to be determined by exact numbers.
These may be signals of integrability of the differential equations, at least in the critical limit.
So far we did not find an analytic proof for the previous findings. 
Another surprising aspect is that the relation $R_{\rm b}=n/(gv)$ between the radius of the bag and the vev seems to be {\it not} renormalized by gravitational corrections (Fig. \ref{radius}).
This was proved analytically using the functional technique.Another important property of the gravitational bags is that $n$ scales out of the equations, and thus the mass per unit of flux is a constant.
This seems to indicate the persistence of the moduli space in the large $n$ limit, but we do not have any progress so far for bags with generic shapes coupled to gravity. 

A final surprise involves the binding energy near the black hole threshold.
We recall that the system in not supersymmetric, and in particular we expect binding energy between multi-monopoles when $\alpha > 0$.
The simplest way to understand this is that since at $\alpha =0$ there is exact cancellation between attractive and repulsive forces, the ones due to the Higgs and the electro-magnetic field respectively; the presence of gravity with $\alpha > 0$ will certainly induce an overall attractive force.
When $\alpha$ reaches the critical value, we instead expect to return to an exact balance, but this time because gravity cancels the electro-magnetic repulsion and the scalar field effect disappears.
In the window $\alpha_{\rm max}^{1} < \alpha < \alpha_{\rm cr}$, where a single monopole is already black hole but the multi-monopole not yet, we can compute the binding energy per unit of flux simply as the deviation of the mass from the critical black hole mass. This vanishes quadratically as $\alpha$ approaches $\alpha_{\rm cr}$.
This mass is the total energy computed from asymptotic infinity. 
If instead we want the binding energy from the point of view of an observer on the bag surface, we have to rescale by the parameter $B(R_{\rm b})$.
This gives a finite amount of binding energy in the $v \to v_{\rm cr}$ limit. 

Many problems are still open. 
The low energy dynamics of magnetic bags has been discussed, as the geometry of an infinite dimensional moduli space.  
But still the  mathematical problem lacks a formal definition and an explicit solution.
Another related issue is to prove (or disprove) the persistence of moduli space of bags for generic $\alpha$. 
Finally it would be good to explain the nice exact relation which we  observed only empirically.

\section*{Acknowledgements}

I thank N. Manton, M. Shifman and D. Tong for useful conversations.
I have presented this topic at DAMTP, FTPI University of Minnesota, and the ``Large N'' workshop at the University of Maryland.
I thank the audiences for interesting questions and comments.
I thank FTPI  and University of Maryland for their hospitality in May 2010.

\end{document}